\documentclass[runningheads]{llncs}
\usepackage{graphicx}
\usepackage{ragged2e}
\usepackage{subfigure} 
\usepackage{paralist}
\begin{document}
\title{Poster: Data Integration for Supporting Biomedical Knowledge Graph Creation at Large-Scale}
%

\author{Samaneh Jozashoori\inst{1,2}\orcidID{0000-0003-1702-8707} \and
Tatiana Novikova\inst{3}\orcidID{0000-0003-1334-0998} \and
Maria-Esther Vidal\inst{2,1}\orcidID{0000-0003-1160-8727}}

\institute{L3S Institute, Leibniz University of Hannover, Germany \\  
\and
TIB Leibniz Information Centre for Science and Technology, Germany
\and
University of Bonn, Germany\\
\email{jozashoori@l3s.de}\\
\email{s6tanovi@uni-bonn.de}\\
\email{maria.vidal@tib.eu}
}
\maketitle              % typeset the header of the contribution

\begin{abstract}
In recent years, following FAIR and open data principles,
the number of available big data including biomedical data has been increased exponentially. In order to extract knowledge, these data should
be curated, integrated, and semantically described. Accordingly, several
semantic integration techniques have been developed; albeit effective,
they may suffer from scalability in terms of different properties of big
data. Even scaled-up approaches may be highly costly because tasks of
semantification, curation and integration are performed independently.
In order to overcome these issues, we devise ConMap, a semantic integration
approach which exploits knowledge encoded in ontology in order to
describe mapping rules to perform these tasks at the same time. Experimental results performed on different data sets suggest that ConMap can significantly reduce the time required for knowledge graph creation by
up to 70\% of the time that is consumed following a traditional approach.
\end{abstract}
\section{Introduction}
\begin{figure}[t!]
\includegraphics[width=\textwidth]{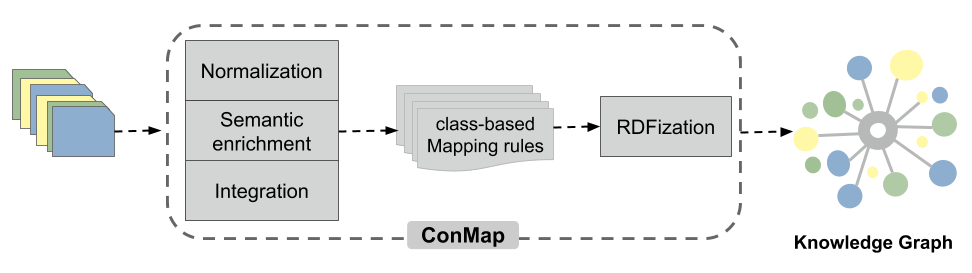}
\caption{\textbf{The ConMap Approach.} ConMap receives structured data sets from heterogeneous sources as input, and produces a knowledge graph. It relies on conceptual or class-based mapping approach in performing all tasks of semantic enrichment, integration, and transformation i.e. both semantification and integration are performed during class-based mapping and afterwards, based on generated mapping rules, normalized data is transformed as RDF model into the knowledge graph.} \label{figure1}
\end{figure}
With the rapid advances in different techniques in biomedical domain such as Next Generation Sequencing \cite{ref_article1}, and access policies such as FAIR \cite{ref_article5} and open data principles, big data has become a quotidian occurrence. However, knowledge discovery from all these big data, as the criteria to make decisions and actions to be taken, is still an unsolved problem. In order to extract knowledge the data should be curated, integrated and semantically described. Recently, the development of Semantic Web technologies has facilitated the implementation of various semantic integration applications, e.g., Karma \cite{ref_site1}, MINTE \cite{ref_article6}, SILK \cite{ref_article9}, and Sieve \cite{ref_article10}. The main purpose of Semantic Web Technologies is to describe the meaning of data, a machine readable fashion. Existing semantic data integration approaches rely on a common framework that allows for transformation of data in various raw formats into a common data model, e.g., RDF \cite{ref_article3}; mapping languages like RML \cite{ref_article2}, are used to express these mappings. Albeit effective, existing semantic data integration tools may suffer from scalability in terms of the dominant dimensions of big data, i.e., volume, variety, veracity, velocity, and value. In fact, even scaled-up approaches mainly scale up in terms of variety, and may be highly
costly since the tasks of semantification, curation, and integration are performed independently.\\
\indent To overcome drawbacks of existing approaches, we introduce ConMap, a semantic integration approach for big data. ConMap exploits knowledge encoded in a global schema to perform all the three mentioned tasks, i.e., semantification, curation, and integration, in a single step; thus, ConMap provides a scalable solution for semantic integration of big data. We have performed initial experiment study over data sets of various sizes; observed results suggest that ConMap reduces the RDFization time \cite{ref_article4}, i.e., the time required for transforming heterogeneous structured data sets into RDF.\\
\indent The rest of the paper is structured as follows: in Section 2 the general idea of ConMap is presented as well as detailed explanation of ConMap architecture and components. In Section 3, the experiments that are performed between ConMap and the attribute-based mapping approach are described and the results are evaluated in terms of time complexity. Finally, Section 4 represents our conclusions.
\section{The ConMap Approach}
ConMap is a semantic data integration approach able to use mapping rules not only for data semantification, but also for curation and integration. ConMap implements a class-based mapping paradigm that resembles the Global-As-View \cite{ref_article7} approach of data integration systems \cite{ref_article8}; it enables the definition of the mapping, curation, and integration rules per each class in the global schema. Thus, ConMap executes all the tasks, i.e., semantification, curation, and integration
at the same time by evaluating these class-based rules. Figure 1 devises the ConMap architecture. ConMap receives real-world data source(s) that represent the same concepts in the global schema but in different formats; it outputs an knowledge graph where input data is integrated and described in a structured way. Data related to each class is extracted from different data sources which are normalized in advance to reduce data duplicates. Afterwards, normalized data is semantified in order to describe and integrate this data in the knowledge graph.
The components of ConMap can be summarized as below:
\begin{itemize}
    \item \textbf{Normalization:} To overcome interoperability issues, all data sets are normalized considering the concepts that are presented. Since each concept may be represented by more than one data source and each data sources may involve more than one concept, the process of normalization is based on the decomposition of each data set in terms of the global schema classes.
    \item \textbf{Semantic Enrichment:} : Based on attributes of each concept that are presented in different data sources, semantic descriptions are added. The classbased mapping approach enables the semantification and curation of entities that are expressed in separated data sources differently.  
    \item \textbf{Integration:} To integrate the data sets derived from normalization of various data sources, semantic descriptions provided during the generation of mapping rules are employed.
    \item \textbf{RDFization:} The last component to be executed in ConMap is RDFization which evaluates class-based mapping rules for transforming normalized and semantified data sets into the knowledge graph.   
\end{itemize}
\begin{figure}[t!]
\includegraphics[width=\textwidth]{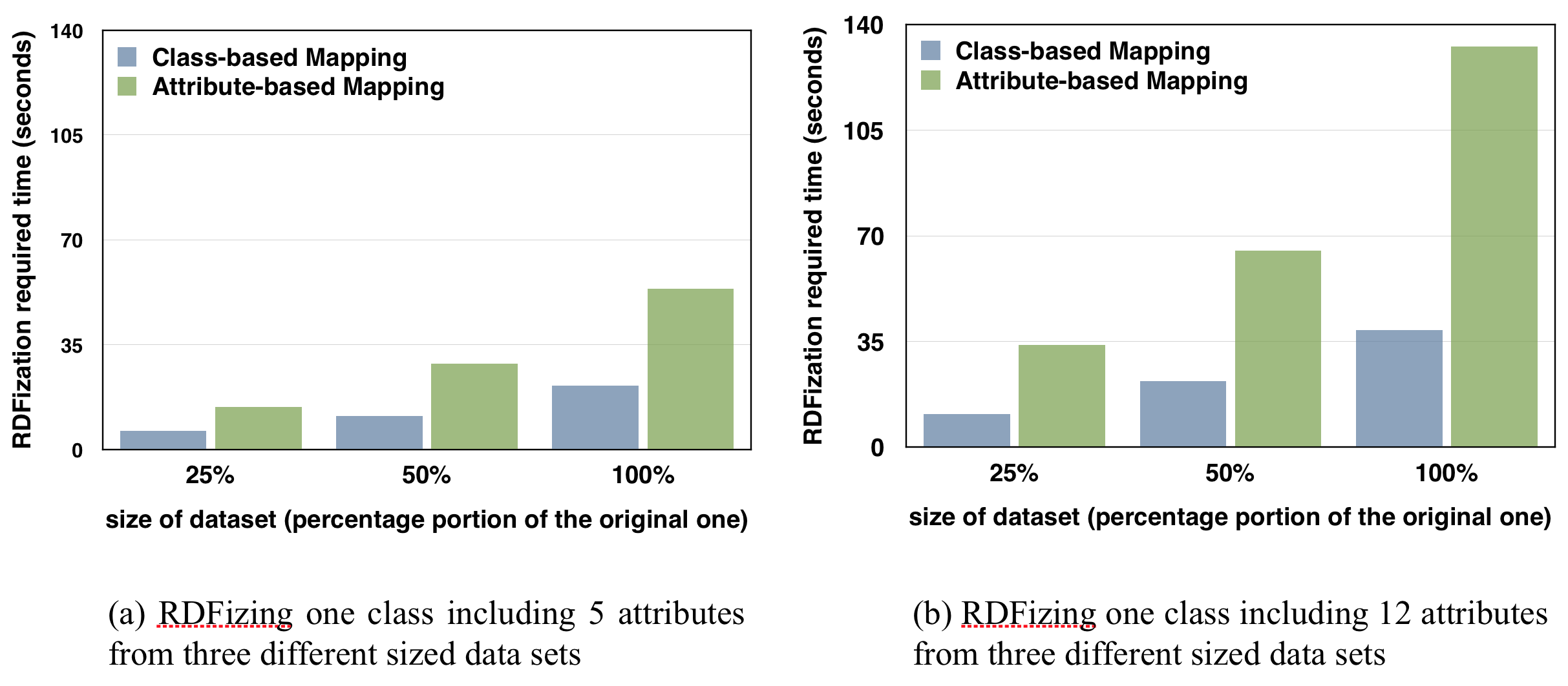}
\caption{\textbf{Experimental results.} (a) The required time for RDFization of one class including five attributes from three different sized data sets. (b) The required time for RDFization of one class including twelve attributes from three different sized data sets.} \label{figure2}
\end{figure}
\section{Experimental Study}
In this paper, the performance of two mapping paradigms are compared: the class-based mapping approach provided by ConMap, and an attribute-based approach which is commonly followed by existing tools, e.g., Karma. We address two research questions:
\begin{inparaenum}[\bf {\bf RQ}1\upshape)]
\item Does ConMap reduce the time complexity of RDFization?
\item How influential a mapping approach can be in terms of execution time when the complexity of the class increases?\\
\end{inparaenum}
\indent \textbf{Benchmark:}
In this study, a data set with overall size of 169.8 MB is extracted from COSMIC\footnote{\url{https://cancer.sanger.ac.uk/cosmic}}, an online database of somatic mutations that are found in human cancer. The data set is in tab separated format comprising 557,162 records of lung cancer related coding point mutations that are derived from targeted and genome wide screens.\\ 
\indent \textbf{Metrics:}
The behavior of the studied mapping approaches is evaluated by measuring the execution time in seconds for transforming a data set into RDF applying that approach.\\
\indent \textbf{Implementation:}
The mapping rules\footnote{https://github.com/samiscoding/DILS} are expressed in the RML mapping language. The RDFization is implemented in Python 3.6. The
experiment was executed on a Ubuntu 17.10 (64 bits) machine with Intel W-2133, CPU 3.6GHz, 1 physical processor; 6 cores; 12 threads, and 64 GB RAM. \\
\indent \textbf{Experimental Setup:}
Two experiments are set up in this study: 
\begin{inparaenum}[\bf {\bf E}1\upshape)]
    \item In order to better understand the influence of mapping approach on time complexity of RDFization, the experiment is run on three different sized data sets: the first one is the preprocessed data set derived from the original mutation data set that is extracted from COSMIC without any decrease regarding its size while the two other data sets are extracted from the first one. The records that are included in two latest data sets are 50\% and 25\% randomly selected records of the first data set. The result of this experiment is shown in Figure \ref{figure2}(a).
    \item To study how time complexity of each mapping approach fluctuated with the increase in the number of attributes for a class, for each mapping approach two separated sets of mapping rules are defined; one mapping rule set for an RDF class with twelve attributes and the other one including five attributes. The experimental results can be seen in Figure \ref{figure2}(b).
\end{inparaenum}
Based on the results of explained experiments that are illustrated in Figure \ref{figure2}, the execution time increases in case of using the attribute-based mapping rules for transformation of data in both sets including different numbers of attributes which positively answers the \begin{inparaenum}[\bf{\bf RQ}1]\item\end{inparaenum}. Moreover, the observed results lead to answer \begin{inparaenum}[\bf {\bf RQ}2]\item \end{inparaenum}as follows: in attribute-based mapping approach, the required execution time for transforming one class of data will grow when the number of its attributes increases, however, in class-based mapping the time complexity is not a function of class complexity. \\
\indent The evaluation results can be simply explained according to the fact that the attribute-based mapping approach performs the same procedure of creating \textit{subject-predicate-object} triple for every single attribute of a class. In contrast, the class-based mapping approach transforms each concept or class including all its attributes to one RDF class in a single run. Therefore, class-based mapping approach can be considered as a fundamental procedure for transforming raw data into RDF model in an integrated non-redundant way. 
\section{Conclusions an Future Work}
We introduced ConMap, a semantic integration approach that deploys the knowledge encoded in an ontology to perform semantification, curation and integration simultaneously, in order to acquire a scale-up integration system. We displayed that ConMap can significantly optimize the transforming of structured data sets to a knowledge graph, in terms of time complexity. Although empirical results demonstrated in this paper were derived by all components of ConMap, there is still to illustrate the power of this approach in terms of integration optimization which will be revealed in future work. 
\section{Acknowledgement}
This work has been supported by the European Union's Horizon 2020 Research and Innovation Program for the project iASiS with grant agreement No 727658. 

\end{document}